\documentclass[conference]{IEEEtran}
\IEEEoverridecommandlockouts
\usepackage{cite}
\usepackage{amsmath,amssymb,amsfonts}
\usepackage{algorithmic}
\usepackage{graphicx}
\usepackage{textcomp}
\usepackage{xcolor}
\usepackage{cite}
\usepackage{amsmath,amssymb,amsfonts}
\usepackage{algorithmic}
\usepackage{graphicx}
\usepackage{textcomp}
\usepackage{xcolor}
\usepackage[hyphens]{url}
\usepackage{hyperref}
\usepackage{booktabs}
\usepackage{multirow}
\usepackage{float}
\usepackage{pifont}
\usepackage[utf8]{inputenc}
\def\BibTeX{{\rm B\kern-.05em{\sc i\kern-.025em b}\kern-.08em
    T\kern-.1667em\lower.7ex\hbox{E}\kern-.125emX}}
\begin{document}

\title{ProtocolLLM: RTL Benchmark for SystemVerilog Code Generation of Communication Protocols\\

}

% \author{\IEEEauthorblockN{Arnav Miteshkumar Sheth}
% \textit{University of Illinois Urbana Champaign}\\
% Champaign, USA \\
% amsheth2@illinois.edu}
% \and
% \IEEEauthorblockN{Ivaxi Sheth}

% \IEEEauthorblockA{\textit{CISPA Helmholtz Center for Information Security}\\
% Saarbruecken, Germany \\
% ivaxi.sheth@cispa.de}
% \and
% \IEEEauthorblockN{Mario Fritz}

% \IEEEauthorblockA{\textit{CISPA Helmholtz Center for Information Security}\\
% City, Country \\
% fritz@cispa.de}

\author{
\IEEEauthorblockN{
Arnav Sheth\IEEEauthorrefmark{1},
Ivaxi Sheth\IEEEauthorrefmark{2},
Mario Fritz\IEEEauthorrefmark{2},
}
%\vspace{0.05in}
\IEEEauthorblockA{
\IEEEauthorrefmark{1}University of Illinois Urbana Champaign. \emph{\href{mailto:amsheth2@illinois.edu}{amsheth2@illinois.edu}}}

\IEEEauthorblockA{\IEEEauthorrefmark{2}CISPA Helmholtz Center for Information Security. \emph{\href{mailto:ivaxi.sheth@cispa.de}{ivaxi.sheth@cispa.de},\href{mailto:fritz@cispa.de}{fritz@cispa.de}}}
%\vspace{0.05in}
}

\maketitle

\begin{abstract}

Recent advances in large language models (LLMs) have demonstrated strong performance in generating code for general-purpose programming languages. However, their potential for hardware description languages (HDLs), such as SystemVerilog, remains largely unexplored. HDL code generation poses unique challenges due to strict timing semantics, concurrency, and synthesizability constraints essential for correct hardware functionality. 
Further, HDL-based design flows encompass a broad set of tasks beyond structural code generation, including testbench development, assertion-based verification, timing closure, and protocol-level integration for on-chip communication.
In this work, we evaluate the capabilities of both open-source and state-of-the-art LLMs in generating synthesizable and functionally accurate SystemVerilog implementations of widely used communication protocols that are critical components of embedded and System-on-Chip (SoC) systems.
We introduce ProtocolLLM, the first benchmark suite specifically targeting these protocols with tasks spanning multiple design abstraction levels and varying prompt specificity. 
Our evaluation method also focuses on timing correctness in addition to synthesizability and syntactic correctness. 
We observe that most of the models fail to generate SystemVerilog code for communication protocols that follow timing constrains\footnote[1]{Code: \url{https://github.com/amsheth/ProtocolLLM} }.

\end{abstract}

\begin{IEEEkeywords}

Large Language Models (LLMs),
hardware description languages,
SystemVerilog,
communication protocols
\end{IEEEkeywords}

\section{Introduction}
\begin{figure}
    \centering
    \includegraphics[width=0.5\textwidth]{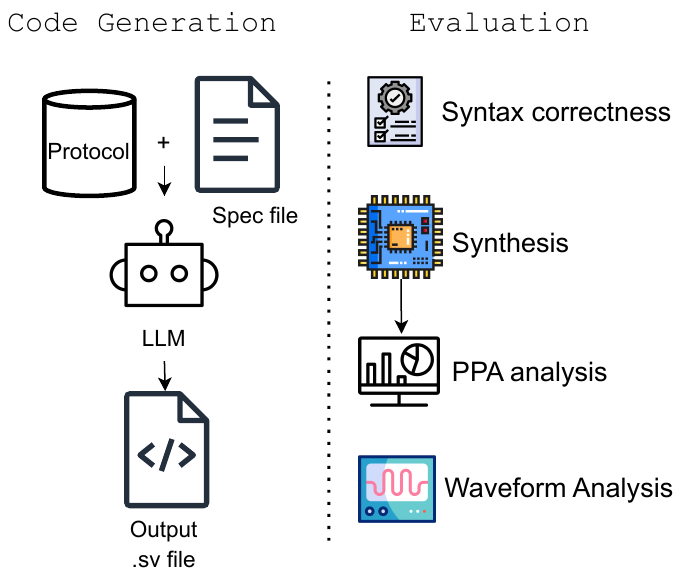}
    \vspace{-4mm}
    \caption{Our HDL code generation and evaluation methodology. The pipeline starts with protocol definitions and optional specification documents as input to an LLM. The generated SystemVerilog code is evaluated through three sequential stages: (1) syntax checking, (2) logic synthesis, and (3) functional simulation. Any failure at these stages leads to rejection, while only code that passes all three is accepted as valid output.}
    \vspace{-5mm}
    \label{teaser}
    % \label{fig:mesh1}
\end{figure}

\begin{table*}[htb!]
\centering
\renewcommand{\arraystretch}{1.2}
\caption{Comparison against similar benchmarks. }
\begin{tabular}{c|c|c|c|c|c|c}
Benchmark & Programming Language & Testcase Types & Syntax & Simulation & Synthesis & PPA \\
\hline
VerilogEval~\cite{liu2023verilogeval} & Verilog & Combinational circuits, FSMs & \ding{51} & \ding{51} & \ding{51} & \ding{55} \\
HDLEval~\cite{10691770} & Multiple & Combinational, pipelined & \ding{51} & \ding{51} & \ding{55} & \ding{55} \\
PyHDL-Eval~\cite{10740201} & Python-embedded DSLs & Spec-based, Verilog, DSL & \ding{51} & \ding{51} & \ding{55} & \ding{55} \\
RTLLM~\cite{RTLLM} & Verilog, VHDL, Chisel & Arithmetic, FSM, Memory, CPU & \ding{51} & \ding{51} & \ding{51} & \ding{51} \\
VHDL-Eval~\cite{vijayaraghavan2024vhdlevalframeworkevaluatinglarge} & VHDL & Translated Verilog, public VHDL & \ding{51} & \ding{51} & \ding{55} & \ding{55} \\
Thakur et al.~\cite{thakur2024verigen} & Verilog & Combinational circuits, FSMs, shift regs & \ding{51} & \ding{51} & \ding{55} & \ding{55} \\
AssertLLM~\cite{Assertllm} & SystemVerilog & Spec-to-Assertion for full modules & \ding{51} & \ding{51} & \ding{55} & \ding{55} \\
Englhardt et al.~\cite{englhardt2023exploringcharacterizinglargelanguage} & C/C++ & Sensor I/O, BLE, I2C & \ding{51} & \ding{51} & N/A & N/A \\
\hline
\textit{ProtocolLLM} & SystemVerilog & Communication protocols - SPI, I2C, UART, AXI & \ding{51} & \ding{51} & \ding{51} & \ding{51} \\
\end{tabular}
\label{tab:comparision}
\end{table*}

Communication protocols are essential to hardware-embedded systems, enabling structured data exchange between processing elements, peripherals, and memory subsystems. In embedded and system-on-chip (SoC) designs, standard protocols such as Serial Peripheral Interface (SPI)~\cite{spi}, Inter-Integrated Circuit (I²C)~\cite{i2c}, Universal Asynchronous Receiver-Transmitter (UART)~\cite{uart}, and Advanced extensible Interface (AXI)~\cite{Arm} are widely used to interface with sensors, actuators, storage devices, and external. The correct and efficient implementation of these protocols in HDL is essential for ensuring functional correctness, timing closure, and system-level integration.

LLMs have shown remarkable performance in code generation, particularly in software domains~\cite{jiang2024survey,jimenez2023swe}. As the complexity of hardware designs continues to grow, there is increasing interest in leveraging language models to assist or automate various stages of the design process~\cite{lu2024rtllm, thakur2024verigen, pan2025survey, RTLLM, Assertllm, chang2023chipgpt, huang2024towards,thakur2023benchmarking,swaroopa2024evaluating}. Their ability to synthesize, complete, and refactor code raises the question: \textit{Can such models generate synthesizable HDL code that adheres to the semantic, structural, and timing constraints of hardware communication protocols?}

Despite the emergence of LLM-based tools for software engineering, their application to hardware domains, especially in generating protocol-level modules, remains limited. 
% Prior work has focused on small-scale or artificial examples, often emphasizing syntax correctness over functional verification or hardware-aware evaluation. Furthermore, existing benchmarks do not reflect the layered complexity of protocol design. 
During hardware development, communication protocols such as SPI, I²C, UART, and AXI serve as fundamental interfaces for inter-module data exchange. These protocols are widely used across various systems, from low-power microcontrollers to high-performance SoCs. Designing synthesizable RTL for these protocols is a non-trivial task that demands strict adherence to timing specifications, finite-state machines (FSMs), signal coordination, and electrical constraints such as clock polarity and phase.

The communication protocol modules are one of the most basic and common examples in VLSI design that also require more holistic evaluation, including waveform-level functional verification, synthesis timing analysis, and deployability on hardware targets such as FPGAs and ASIC's.
In particular, protocol designs involve multi-signal interactions, such as Ready/Busy/ACK Signals, MOSI/MISO, and Multibyte data transmission, which must adhere to strict temporal relationships, making surface-level correctness insufficient for meaningful evaluation.

Current HDL code generation works have largely focused on isolated or synthetic examples in Verilog and VHDL~\cite{thakur2024verigen, liu2023verilogeval, yubeaton2025verithoughts, wei2025vflow,garcia2025turtle,li2025autosilicon}. Many of them focus on syntax correctness alone~\cite {RTLLM, Assertllm, chang2023chipgpt} for evaluation. In contrast, protocol-level designs such as SPI, I²C, or AXI demand adherence to precise timing relationships, signal-level interactions, and behavioral specifications derived directly from detailed datasheets. These implementations require waveform-accurate behavior and rigorous validation against temporal constraints, which are not captured by syntax-based benchmarks.

Given the adoption of SystemVerilog across the semiconductor industry for both hardware design and verification, our task requires code generation specifically in SystemVerilog.
Unlike traditional Verilog benchmarks, SystemVerilog's advanced constructs directly facilitate the modeling of complex protocols and interactions. 

% Unlike traditional Verilog benchmarks, SystemVerilog's advanced constructs directly facilitate modeling complex protocols and interactions.

% SystemVerilog is an extended version of Verilog that includes additional features for design and verification, particularly for large and complex designs. While Verilog focuses on hardware description, SystemVerilog provides features for both design and verification, including object-oriented programming, advanced data types, and enhanced verification constructs.

 % We are also one of the first benchmarks to implement the study in System Verilog. 

% ystemVerilog is crucial for accurately modeling hardware behavior due to its ability to combine precise hardware description with sophisticated verification features. This allows for waveform-level accuracy and adherence to stringent timing constraints.  We are also one of the first benchmarks to implement this study in SystemVerilog.

We introduce the first benchmark in System Verilog for evaluating LLMs on HDL-based communication protocol generation to address this gap (See \autoref{teaser}). Our benchmark covers multiple protocols: SPI, I²C, UART, and AXI. The benchmark focuses on open-ended code generation, where models are required to synthesize complete, synthesizable modules that meet protocol-level functional and temporal constraints.

In our open-source benchmark, we perform extensive experiments across different LLMs, including code models and general-purpose models. The code is comprehensively evaluated in four stages: (1) Lint to ensure language-level correctness and synthesizability, (2) Logic Synthesis using open source EDA tools to assess post-silicon quality, (3) Waveform Analysis to validate temporal behavior against golden references. 
Finally, our contributions can be summarised below:
\begin{enumerate}
    \item We propose the \textbf{first} structured benchmark focused on communication protocol generation using LLMs, spanning widely used industry protocols: SPI, I²C, UART, and AXI. 

    \item We introduce a three-stage evaluation framework tailored to hardware development workflows. This includes (i) lint checks, (ii) logic synthesis, and (iii) waveform validation for assessing hardware resource usage, maximum achievable frequency, and area overhead.

    \item We evaluate a diverse set of LLMs, including both code fine-tuned models and general-purpose models under both vanilla and retrieval-augmented generation (RAG).
\end{enumerate}

\section{Background}

\subsection{HDL Code Generation with LLM}

Recent works have revealed weaknesses of LLMs for HDL generation, particularly in synthesis compatibility, functional correctness, and verification logic quality. Benchmarks like RTLLM~\cite{RTLLM} and VerilogEval~\cite{liu2023verilogeval} show that LLMs frequently generate non-synthesizable HDL due to incorrect timing constructs, vendor-incompatible syntax, and structural flaws ~\cite{RTLLM, Assertllm, chang2023chipgpt, huang2024towards,thakur2023benchmarking,swaroopa2024evaluating}. Formal tools such as Cadence Jasper have further identified that up to $60\%$ of LLM-generated designs contain critical weaknesses like incorrect state transitions and bit-width mismatches \cite{gadde2024all, jha2024automated}, while iterative techniques such as counterexample-guided refinement offer only partial mitigation \cite{jha2024automated, huang2024towards}. Despite syntax improvements, logical correctness remains a persistent issue, with misconfigured ports, flawed carry logic, and incomplete FSM implementations reported by previous works. 
% Verification logic, including testbenches and SystemVerilog Assertions (SVAs), often lacks coverage and adherence to industry practices like UVM \cite{pei2024betterv}, highlighting insufficient feedback integration.

Unlike prior benchmarks (see~\autoref{tab:comparision}) that focus on isolated RTL modules or general-purpose logic generation, our work targets protocol-specific HDL generation, incorporating temporal \textbf{and} functional constraints derived from real-world datasheets. Furthermore, we introduce a multi-stage evaluation pipeline, including waveform-level validation, which more closely aligns with practical hardware verification flows.

\subsection{Communication Protocol}

A common and critical application of FPGAs and microcontrollers is the implementation and management of communication protocols. These protocols are essential for interfacing components such as sensors~\cite{kumari2017interfacing}, actuators~\cite{bosse2010smart}, and master controllers in embedded systems~\cite{subero2024usart}. In this benchmark, we focus on four widely used inter- and intra-system communication protocols: SPI, I²C, UART, and AXI.

\begin{itemize}
    \item \textbf{SPI}~\cite{spi} is a synchronous serial communication interface that employs a master-slave architecture, where a single master device orchestrates communication with one or more slave devices. The master device generates the clock signal on the SCLK line, which synchronizes data transmission between the master and the selected slave. Unlike I2C's addressing scheme, SPI often uses a dedicated Slave Select (SS) line for each slave device. The master activates a specific slave by pulling its SS line low before initiating data transfer. Data is simultaneously transmitted and received between the master and the slave on the MOSI and MISO lines, respectively, during each clock cycle.
    \item \textbf{I2C}~\cite{i2c} is a synchronous, multi-master/multi-slave serial communication bus that utilizes only two bidirectional open-drain lines: the Serial Data Line (SDA) and the Serial Clock Line (SCL). I2C employs a master-slave architecture where one or more master devices initiate and control communication with multiple slave devices. The master device is responsible for generating the clock signal on the SCL line that synchronizes all data transfers. Each slave device connected to the I2C bus is assigned a unique address, which the master uses to select and communicate with a specific slave. This addressing scheme allows multiple slave devices to share the same two communication wires, a significant advantage over SPI, which typically requires a dedicated select line for each slave. 

    \item \textbf{UART}~\cite{uart} is an asynchronous, full-duplex serial communication protocol commonly employed for point-to-point data exchange. Unlike synchronous protocols, UART does not require a shared clock signal instead, its baud rate must be configured identically on both communicating devices. Data is transmitted serially, one bit at a time, over two dedicated lines: the Transmit (Tx) line and the Receive (Rx) line. Communication is packet-based, with each data frame consisting of a start bit, a configurable number of data bits (typically 8), an optional parity bit for error detection, and one or more stop bits. 
    \item \textbf{AXI4 (Lite)}~\cite{Arm} is an interface protocol defined by ARM as part of the AMBA (Advanced Microcontroller Bus Architecture) standard.
    % It is one of the most common Communication Bus Protocol. It is commonly used to connect a FPGA and external cores. The AXI protocol defines the interaction between master devices, which initiate communication transactions, and slave devices, which respond to these transactions. 
    While AXI is inherently a point-to-point interface, its architecture can be extended to support multiple masters and slaves through the use of interconnect components, which act as intelligent switches and routers on the chip. The protocol mandates a handshake-like procedure involving VALID and READY signals for each transmission, ensuring reliable data transfer by synchronizing the data flow between the source and destination. AXI-Lite bus is a smaller version of AXI bus that only supports a single ID thread per initiator.
\end{itemize}
These protocols were selected based on their practical relevance and increasing complexity, enabling a gradient of difficulty for evaluating the ability of LLM to synthesize code.

\section{ProtocolLLM}

In this paper, we present a comprehensive benchmark - \emph{ProcotcolLLM} designed to evaluate LLMs for the task of generating synthesizable HDL implementations of communication protocols. These protocols serve as foundational components in modern hardware systems, facilitating the communication between processing units, peripherals, and memory subsystems. The complexity of accurately implementing these protocols in HDL, particularly as SystemVerilog, requires strict adherence to timing constraints, signal coordination, and FSM behavior, making it an ideal domain for testing LLM-based code generation. 

Our benchmark is designed to evaluate both the functional and timing correctness of LLM-generated designs across these protocols.

\subsection{Design Principles}

The design of our benchmark is guided by the following core principles:

\begin{itemize}
    \item \textbf{Protocol-Centric Evaluation}: The benchmark focuses on a set of widely-used communication protocols: SPI, I²C, UART, and AXI that are essential for interfacing between processing elements and peripherals. These protocols represent different levels of complexity in terms of signaling, timing, and interaction, providing a diverse set of challenges for model evaluation.
    
    \item \textbf{Synthesizability and Timing Fidelity}: Beyond syntax, the generated HDL must conform to hardware implementation constraints, including FSM integrity, signal synchronization, and timing closure. This ensures that models produce RTL code that is not only valid but also deployable on real-world FPGA and ASIC hardware.

    \item \textbf{Specification-Guided Generation}: Instead of synthetic or abstract prompts, our tasks are grounded in protocol specification documents and datasheets. This setting reflects practical design conditions where engineers synthesize HDL directly from interface specifications, enabling a more realistic assessment of model utility.

    \item \textbf{Protocol Variations} We introduced targeted variations within each protocol to assess the LLMs’ ability to capture fundamental design requirements. For SPI, this involved testing all four CPOL/CPHA mode combinations; for I²C, evaluating both 7-bit and 10-bit addressing schemes; and for UART, comparing transmission with and without parity. 

    \item \textbf{Holistic Evaluation Pipeline}: Our benchmark includes a three-stage evaluation framework encompassing (i) linting for syntax and synthesis readiness, (ii) logic synthesis for area, frequency, and utilization metrics, and (iii) waveform analysis for functional validation against protocol timing diagrams and expected behaviors.
\end{itemize}

\subsection{Benchmark Tasks and Evaluation Criteria}

Our primary task is open-ended full module generation, where the model receives a natural language or formal protocol description and is tasked with generating a complete, synthesizable SystemVerilog module. This evaluates the model's ability to interpret high-level requirements and produce corresponding RTL code.

\begin{figure}[h]
    \centering
    \vspace{-3mm}
\includegraphics[width=0.45\textwidth]{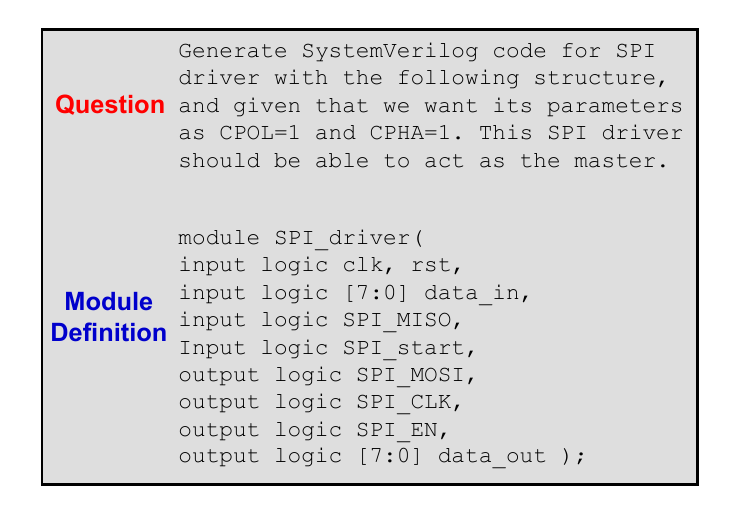}
    \vspace{-5mm}
    \caption{Sample prompt for SPI protocol.}
    \vspace{-1mm}
    \label{fig:mesh1}
\end{figure}

We prompt the model in two ways:

\begin{itemize} \item \textbf{Standard Prompting}: The model generates HDL code based on a full problem description and module specification (see \autoref{fig:mesh1}). 
\item \textbf{Spec-Assisted Generation}: In addition to the standard prompting, the model is assisted via retrieval-augmented generation (RAG). Here, relevant parts from the document are retrieved and appended to the prompt to potentially improve the prior and accuracy.
\end{itemize}

Evaluation of a code generated by LLM is challenging, as LLMs are usually trained to write code for object-oriented programming languages like C/C++ and Python, which are usually not time-dependent and do not generate combinational and sequential logic. Our criteria to evaluate SystemVerilog code are on three-prong basis to wholistically evaluate the generated code. 
The criteria in the order are as follows:

% from typical specification documents and datasheets.     

\begin{itemize}
    \item \textbf{Lint Check}: We use \textbf{Verilator} Lint, an open-source linting tool, as the first step in the evaluation pipeline. This tool scans the generated SystemVerilog code for syntactical issues, undeclared signals, improper assignments, and other potential logical errors that may prevent successful simulation or synthesis. Catching these issues early ensures that only structurally correct and semantically valid designs proceed to simulation and synthesis, reducing the chances of failure in downstream verification or implementation stages.

% waive -rule {{WarnAnalyzeBBox}}  -comment {Design Unit '' does not have functional view in any of the given sglib files}
% waive -rule {{NoAssignX-ML}}     -comment {RHS of a assignment contains 'X}
% waive -rule {{STARC05-2.11.3.1}} -comment {FSM must be split into always_ff and always_comb}
% waive -rule {{FlopEConst}}       -comment {Flip-flop enable pin is permanently disabled or enabled}
% waive -rule {{W528}}             -comment {Variable set but not read}
% waive -rule {{STARC05-2.2.3.3}}  -comment {multiple assignment in always_ff}
% waive -rule {{W287b}}            -comment {output of a instance not connected}
% waive -rule {{W415a}}            -comment {multiple assignment in always_comb}
% waive -rule {{ErrorAnalyzeBBox}} -comment {Design Unit '%s' has no definition; black-box behavior assumed and module interface inferred}
% waive -rule {{SYNTH_5166}}       -comment {Non-synthesizable statement ignored}

    \item \textbf{Synthesis}: Synthesis analysis is performed using \textbf{Yosys}, an open-source synthesis tool. This step evaluates whether the generated SystemVerilog code is structurally and semantically valid for hardware realization. It provides insights into resource utilization, such as gate count and area, and determines whether the design adheres to the constraints necessary for downstream implementation.

    \begin{itemize}
         \item  \textbf{Power and Area}: \textbf{Yosys} and \textbf{OpenSTA} are used to perform timing, power, and area analysis on the synthesized designs. The clock period is conservatively set to 10 ns, allowing generous timing slack to identify functional rather than performance-related issues. Area estimation is derived from the gate-level netlist using the \textbf{OSU018 standard cell} library during Yosys synthesis. For power analysis, we extract static power consumption using reports generated by OpenSTA, providing insights into the baseline power characteristics of the synthesized designs under no switching activity.
    \end{itemize}

    \item \textbf{Waveform Analysis}: The waveform analysis will determine whether the generated code correctly implements the communication protocol. We simulate the generated modules using \textbf{UVM}-based testbenches that provide a structured and reusable environment for verification. These testbenches incorporate randomized stimulus generation, protocol-specific sequence items, and scoreboards to compare expected and actual behavior. By injecting randomized vectors and monitoring signal activity, we evaluate whether the module correctly transmits or receives data according to the protocol specification. The resulting waveforms are then inspected to confirm compliance with timing, handshake behavior, and protocol semantics.

    % \item \textbf{PPA} \textcolor{red}{FINISH this!}
\end{itemize}

To score the different code generations,  we use Pass@K metric, which measures the probability that at least one of the $k$ generated outputs satisfies the specified test criteria. We set $k=1$ to reflect a realistic deployment scenario such as Co-pilot, where the model is expected to produce a functionally correct implementation on the first attempt following \cite{garcia2025turtle}.
Finally, the Lint, Synthesis, and Waveform results are averaged across all prompt variations for each protocol, providing a general assessment of how each LLM performs under different variations of the protocol.

\begin{table*}[h]
\centering
\caption{Evaluation Results for HDL Communication Protocol Generation. We report Pass@1 results when prompted with and without specification files.}
\begin{tabular}{ll|cc|cc|cc|cc|cc}
\toprule
\textbf{Model} & \textbf{Protocol} 
& \multicolumn{2}{c|}{\textbf{Lint Pass}} 
& \multicolumn{2}{c|}{\textbf{Synthesis}} 
& \multicolumn{2}{c|}{\textbf{Waveform}}
& \multicolumn{2}{c|}{\textbf{Power(mW) }}
& \multicolumn{2}{c}{\textbf{Area (µm²)}} \\
\cmidrule(lr){3-4} \cmidrule(lr){5-6} \cmidrule(lr){7-8} \cmidrule(lr){9-10} \cmidrule(lr){11-12} 
& & \textbf{w/o Spec} & \textbf{w Spec} 
  & \textbf{w/o Spec} & \textbf{w Spec}
  & \textbf{w/o Spec} & \textbf{w Spec} 
  & \textbf{w/o Spec} & \textbf{w Spec} 
  & \textbf{w/o Spec} & \textbf{w Spec} \\
\midrule

\multirow{4}{*}{Deepcoder 14B} 
& SPI  & 0.25 & 0.5 & 0 & 0.25 & 0 & 0 & 0 & 9.29 & 0 & 27184\\
& I²C  & 0 & 0 & 0 & 0 & 0 & 0 & 0 & 0 & 0 & 0\\
& UART & 0 & 0 & 0 & 0 & 0 & 0 & 0 & 0 & 0 & 0\\
& AXI  & 0 & 0 & 0 & 0 & 0 & 0 & 0 & 0 & 0 & 0\\

\midrule

\multirow{4}{*}{QwenC 2.5-14B} 
& SPI  & 1 & 1 & 0.5 & 0.5 & 0 & 0 & 4.52 & 6.19 & 23336 & 21884\\
& I²C  & 0 & 1 & 0 & 0 & 0 & 0 & 0 & 0 & 0 & 0 \\
& UART & 0 & 0 & 0 & 0 & 0 & 0 & 0 & 0 & 0 & 0\\
& AXI  & 0 & 0 & 0 & 0 & 0 & 0 & 0 & 0 & 0 & 0\\

\midrule

\multirow{4}{*}{QwenC 2.5-32B} 
& SPI  & 0.75 & 0.75 & 0.75 & 0.25 & 0 & 0 & 4.46 & 2.28 & 17008 & 27208\\
& I²C  & 0 & 0 & 0 & 0 & 0 & 0 & 0 & 0 & 0 & 0\\
& UART & 0 & 0 & 0 & 0 & 0 & 0 & 0 & 0 & 0 & 0\\
& AXI  & 0 & 1 & 0 & 1 & 0 & 0 & 0 & 52.0 & 0 & 118284\\

\midrule

\multirow{4}{*}{Claude Sonnet 4} 
& SPI  & 1 & 1 & 1 & 1 & 0 & 0 & 7.53 & 6.79 & 40254 & 35178\\
& I²C  & 1 & 1 & 1 & 1 & 0 & 0 & 4.62 & 4.93 & 63424 & 73522\\
& UART & 0.5 & 1 & 0.5 & 1 & 0 & 0 & 14.4 & 6.38 & 96932 & 78264\\
& AXI  & 1 & 1 & 1 & 1 & 0 & 0 & 59.1 & 63.2 & 111884 & 116660\\

\midrule

\multirow{4}{*}{Claude-opus} 
& SPI  & 1 & 1 & 1 & 1 & 0.25 & 0.25 & 11.7 & 9.55 & 36480 & 40668 \\
& I²C  & 1 & 1 & 1 & 1 & 0 & 0 & 11.8 & 12.9 & 62256 & 58804\\
& UART & 0.5 & 1 & 0.5 & 1 & 0 & 0.5 & 6.31 & 12.1 & 97168 & 82368\\
& AXI  & 1 & 1 & 1 & 1 & 0 & 0 & 66.2 & 116 & 122484 & 202684 \\

\midrule

\multirow{4}{*}{Gemini 2.5 Flash} 
& SPI  & 1 & 0.5 & 1 & 0.5 & 0.5 & 0 & 10.4 & 9.63 & 37297 & 31822\\
& I²C  & 0.5 & 0.5 & 0 & 0.5 & 0 & 0 & 0 & 18.5 & 0 & 80896 \\
& UART & 0 & 0 & 0 & 0 & 0 & 0 & 0 & 0 & 0 & 0\\
& AXI  & 1 & 1 & 1 & 1 & 0 & 0 & 60.7 & 16.3 & 145728 & 38000 \\

\midrule

\multirow{4}{*}{Gemini 2.5 Pro} 
& SPI  & 0.75 & 0.5 & 0.75 & 0.5 & 0.5 & 0.5 & 12.5 & 9.93 & 36493 & 29978 \\
& I²C  & 0.5 & 0 & 0.5 & 0 & 0 & 0 & 50.7 & 0 & 84384 & 0\\
& UART & 0.5 & 0.5 & 0 & 0.5 & 0 & 0 & 0 & 0.77 & 0 & 11508\\
& AXI  & 1 & 1 & 1 & 1 & 0 & 1 & 60.7 & 60 & 112316 & 111820 \\

\midrule

\multirow{4}{*}{GPT o3} 
& SPI  & 1 & 1 & 0.75 & 0.5 & 0.25 & 0 & 11.3 & 10.9 & 37545 & 32894 \\
& I²C  & 1 & 1 & 0.5 & 0 & 0 & 0 & 1.41 & 0 & 25548 & 0 \\
& UART & 0.5 & 0.5 & 0.5 & 0.5 & 0 & 0.5 & 0 & 8.86 & 0 & 133768 \\
& AXI  & 1 & 1 & 1 & 1 & 1 & 0 & 72.8 & 60.9 & 111564 & 111832 \\

\midrule

\multirow{4}{*}{GPT 4o} 
& SPI  & 0.75 & 0.75 & 0.75 & 0.75 & 0 & 0 & 11.6 & 5.15 & 27122 & 13884\\
& I²C  & 0 & 0 & 0 & 0 & 0 & 0 & 0 & 0 & 0 & 0\\
& UART & 1 & 0.5 & 0 & 0.5 & 0 & 0 & 0 & 7.09 & 0 & 66708\\
& AXI  & 1 & 0 & 1 & 0 & 0 & 0 & 13.5 & 0 & 37496 & 0 \\

\midrule

\multirow{4}{*}{GPT-4.1} 
& SPI  & 1 & 1 & 1 & 1 & 0.75 & 0.75 & 9.61 & 11.0 & 28729 & 35153\\
& I²C  & 0.5 & 0 & 0.5 & 0 & 0 & 0 & 3.63 & 0 & 55016 & 0 \\
& UART & 1 & 0 & 1 & 0 & 1 & 0 & 16.1 & 0 & 86448 & 0\\
& AXI  & 1 & 1 & 1 & 1 & 0 & 0 & 64.8 & 66.2 & 124552 & 123316\\

\bottomrule
\end{tabular}
\label{main}
\end{table*}

\subsection{Experimental Setup}
% We evaluate the benchmark on open-source coder models: Qwen Coder 2.5-14B and Qwen Coder 2.5-32B ~\cite{hui2024qwen2}. We also evaluate on a general-purpose proprietary GPT-4.1 model~\cite{achiam2023gpt}. All models are queried with a temperature of 0 to ensure deterministic outputs for reproducibility.

We evaluate the benchmark on a diverse set of open-source and proprietary LLMs. Among the open-source models, we include QwenCoder 2.5–14B~\cite{hui2024qwen2}, QwenCoder 2.5–32B~\cite{hui2024qwen2}, Deepcoder 14B~\cite{huang2024opencoder}. We also test on leading closed general purpose models: Claude Sonnet 4~\cite{claudesonnet}, as well as Claude-Opus~\cite{claudeopus}, Gemini 2.5 Pro~\cite{comanici2025gemini25pushingfrontier}, Gemini 2.5 Flash~\cite{comanici2025gemini25pushingfrontier}, GPT-o3, GPT-4o~\cite{openai2024gpt4technicalreport} and GPT-4.1~\cite{achiam2023gpt}. All models are queried with a temperature of 0 to ensure deterministic and reproducible outputs across all HDL protocol generation tasks, allowing fair and consistent evaluation of functional correctness, synthesis viability, and waveform accuracy.

\section{Results}

Our benchmark evaluation in \autoref{main} reveals several important patterns about the capabilities and limitations of current LLMs in HDL protocol generation, particularly when targeting synthesizable and functionally correct modules.

\textbf{Overall Results.} At a system-wide level, bigger models like GPT 4.1 and Gemini 2.5 Pro tend to outperform all smaller models across all evaluation axes: syntax, synthesis, and waveform fidelity. 
We notice that GPT 4.1 has the highest pass rate. 

By incorporating comprehensive evaluation across syntax (linting), synthesis, and functional simulation, we observe a consistent pattern that while the code generated is structurally or syntactically valid HDL, they often fail to meet the synthesizability requirements. These failures include inferred latches, incomplete state transitions, and timing violations that are not detectable through syntax or synthesis checks alone. This shows the significance of including waveform-level simulation in our evaluation pipeline and general HDL benchmarks. It could serve as the final and rigorous validation step, revealing subtle protocol-level misbehaviors and logic flaws that earlier stages cannot capture.

\textbf{Differences across protocols.} Interestingly, we observe a clear protocol-specific performance bias: simpler serial protocols like SPI and AXI4 Lite were more reliably generated, while UART and I2C, which involve greater temporal complexity and signal interplay, exposed deeper model weaknesses.

\textbf{RAG.} The implementation of Retrieval-Augmented Generation that gives relevant spec document actually yielded mixed results. While the number of syntactically correct and synthesizable outputs was slightly lower compared to non-RAG prompts, RAG often led to improved functional correctness as observed in waveform-level simulations. This suggests that incorporating specification documents into the prompt context helps LLMs better understand the behavior and structure of communication protocols. However, the increased context length introduced by RAG may also lead to semantic degradation, where models produce code that is less consistent or structurally flawed despite capturing higher-level intent more accurately, especially for open code models.

\textbf{Power and Area Analysis.} It was observed that the SystemVerilog code generated by LLM that required significantly abnormal power consumption (compared to other implementations) had a higher tendency to fail waveform simulation. This suggests that extreme synthesis characteristics, such as unusually large area or abnormal power profiles, may reflect underlying logic errors or suboptimal code structures. Further, comparing all functionally correct SPI implementations, we observe that GPT-4.1 without RAG produced the most area and power efficient design. In contrast, Claude Opus exhibited the highest area footprint, consuming approximately 1.5× more area, while Gemini 2.5 Pro had the highest power consumption, drawing roughly 1.3× more static power than GPT-4.1. These findings highlight the potential for code-level optimizations even among functionally correct designs and underscore the need to evaluate LLM-generated hardware not only for correctness, but also for implementation efficiency.

\textbf{Effect of code finetuning.} Coder models are typically fine-tuned on large-scale code-specific datasets~\cite{huang2024opencoder, li2023starcoder, lozhkov2024starcoder}, which often contain limited or no HDL, particularly SystemVerilog. This leads to a domain mismatch and out-of-distribution behavior when these models are applied to hardware design tasks. A notable qualitative observation, we made with DeepCoder-14B, which generated Python code instead of SystemVerilog for I²C and SPI protocol prompts. However, this issue was partially mitigated when the model was provided with spec, where the output was SystemVerilog code. Nonetheless, most of the generated code still failed to pass even basic syntax checks.
This behavior suggests that strong priors from fine-tuning on general-purpose software languages can bias the model away from hardware-specific syntax and semantics. Our result on low performance with code models is also supported by recent works where general pupose models perform better than code models~\cite{garcia2025turtle}.
This highlights the need for dedicated models or domain-specific finetuning for HDL and hardware-related generation tasks.
% This highlights a significant limitation in the model’s modality adherence and suggests a lack of robustness in following structured output constraints, especially for domain-specific hardware description tasks.

% Semantic conditioning via spec files had nuanced effects. For simpler protocols like SPI, spec guidance significantly improved synthesis success, e.g., QwenCoder2.5-14B went from full fail to passing synthesis.

% As a side note, what we consider warnings are basically what a good and ideal code should not contain, but are both syntactically correct and do not affect its ability. For example, a signed to unsigned conversion, ports being declared but not read, asynchronous reset, and write-write race for signals can create a functional but not well-developed code. Warnings are just a preference of what good System Verilog code should be, however, they do not affect the code's ability to run a simulation. 

\textbf{Near-miss phenomenon.} LLMs often generated code with minor syntax yet critical functionality issues, such as failing to properly assert or de assert handshake signals, missing initial reads, or leaving ready or ack signals high indefinitely. While these errors do not always prevent synthesis, they often lead to functional failures during simulation or waveform analysis. This highlights the potential value of a “LLM + post-processing” strategy, where automated lint-fixing, formal verification, or human-in-the-loop design repair can be employed to correct subtle flaws. 
% Such approaches can particularly promising when combined with specification-aware prompting, as the model’s output tends to be closer to the correct structural intent, requiring only minimal guided correction to achieve functional correctness. 

\textbf{Variation to subtle changes.} Interestingly, varying Clock Polarity (CPOL) and Clock Phase (CPHA) configurations in SPI, though semantically equivalent in behavior, often resulted in structurally different implementations across model generations. In some cases, models would generate a monolithic FSM for one configuration while producing a modular or decomposed FSM for another. This inconsistency reflects a lack of canonical understanding of protocol design patterns and suggests that models may struggle to generalize protocol semantics across parameterized variants. Such variability in structural representation can hinder downstream reproducibility and verification, especially in hardware design workflows that rely on predictable and uniform architecture.
\section{Conclusion}

% We presented ProtocolLLM the first benchmark to evaluate LLM on System Verilog protocol generation. 

% We presented the first benchmark designed to evaluate large language models on HDL-based communication protocol generation. Our results show that while larger models like QwenCoder2.5-32B and GPT-4.1 produce syntactically correct SystemVerilog, they often fail to meet functional and synthesis-level correctness. The introduction of specification-aware prompting improved both the structure and utility of generated designs, particularly for well-documented protocols like SPI and I²C. However, complex protocols such as AXI remained challenging across all models. Our evaluation pipeline spanning linting, synthesis, and waveform simulation seemed to be critical for identifying errors that surface only in post-synthesis or functional testing. These findings highlight the current limitations of LLMs in hardware design and point to the need for more domain-specific tuning and model refinement.

% In this work, we introduced ProtocolLLM, a comprehensive benchmark designed to systematically evaluate Large Language Models (LLMs) in generating synthesizable and functionally correct SystemVerilog implementations of standard communication protocols (SPI, I²C, UART, and AXI). Our multi-stage evaluation approach—comprising lint checks, synthesis viability, waveform-level simulation with UVM testbenches, and power-area analysis—provided detailed insights into model performance beyond syntax correctness alone.

Hardware description languages (HDLs) like SystemVerilog play a critical role in chip design. Despite the popularity of code-assistants, the application of large language models (LLMs) to HDL generation remains underexplored and faces unique challenges. These include syntax correctness, logic synthesis—including power, performance, and area (PPA) considerations and more importantly, timing constrains. To focus on these, we introduce ProtocolLLM, the first benchmark specifically for SystemVerilog code and the first to focus on communication protocol implementations. Through multi-stage evaluation, our results show that syntax-level accuracy is insufficient for assessing HDL code generation and current LLMs lack robustness in handling the strict timing and modularity constraints of hardware design. Overall, general purpose models such as GPT-4.1 and Gemini 2.5 Pro showed strongest performance in comparision on code models, that were potentially finetunened on large software data. We hope that our benchmark plays a pivotal role in advancing research toward more domain-aware LLMs for hardware design automation with timing constrains.

\section{Future Work}
Our benchmark covers four key communication protocols: SPI, I²C, UART, and AXI, that can be extended to include a broader range of industrial HDL designs such as bus arbiters, memory controllers, and custom pipelines. Future work may also incorporate integration- and system-level behaviors like protocol interoperability and backpressure handling.
Additionally, many bus protocols implement hardware security features, including privilege enforcement and address protection. Expanding the benchmark to evaluate these security aspects will provide deeper insights into HDL generation quality.

\section*{Acknowledgments}
This work was partially supported by the ELLIOT Grant funded by the European Union under grant agreement No. 101214398.

\bibliographystyle{IEEEtranS}
\bibliography{refs}

% Generated by IEEEtranS.bst, version: 1.14 (2015/08/26)
\begin{thebibliography}{10}
\providecommand{\url}[1]{#1}
\csname url@samestyle\endcsname
\providecommand{\newblock}{\relax}
\providecommand{\bibinfo}[2]{#2}
\providecommand{\BIBentrySTDinterwordspacing}{\spaceskip=0pt\relax}
\providecommand{\BIBentryALTinterwordstretchfactor}{4}
\providecommand{\BIBentryALTinterwordspacing}{\spaceskip=\fontdimen2\font plus
\BIBentryALTinterwordstretchfactor\fontdimen3\font minus \fontdimen4\font\relax}
\providecommand{\BIBforeignlanguage}[2]{{%
\expandafter\ifx\csname l@#1\endcsname\relax
\typeout{** WARNING: IEEEtranS.bst: No hyphenation pattern has been}%
\typeout{** loaded for the language `#1'. Using the pattern for}%
\typeout{** the default language instead.}%
\else
\language=\csname l@#1\endcsname
\fi
#2}}
\providecommand{\BIBdecl}{\relax}
\BIBdecl

\bibitem{claudeopus}
\BIBentryALTinterwordspacing
``Claude opus.'' [Online]. Available: \url{https://www.anthropic.com/claude/opus}
\BIBentrySTDinterwordspacing

\bibitem{claudesonnet}
\BIBentryALTinterwordspacing
``Claude sonnet.'' [Online]. Available: \url{https://www.anthropic.com/claude/sonnet}
\BIBentrySTDinterwordspacing

\bibitem{Arm}
ARM, \emph{\BIBforeignlanguage{English}{Introduction to AMBA AXI4}}, ARM limited, 2020.

\bibitem{10740201}
C.~Batten, N.~Pinckney, M.~Liu, H.~Ren, and B.~Khailany, ``Pyhdl-eval: An llm evaluation framework for hardware design using python-embedded dsls,'' in \emph{2024 ACM/IEEE 6th Symposium on Machine Learning for CAD (MLCAD)}, 2024, pp. 1--17.

\bibitem{bosse2010smart}
S.~Bosse and D.~Lehmhus, ``Smart communication in a wired sensor-and actuator-network of a modular robot actuator system using a hop-protocol with delta-routing,'' in \emph{Proceedings of smart systems integration conference, Como, Italy}, 2010, pp. 23--24.

\bibitem{chang2023chipgpt}
K.~Chang, Y.~Wang, H.~Ren, M.~Wang, S.~Liang, Y.~Han, H.~Li, and X.~Li, ``Chipgpt: How far are we from natural language hardware design,'' \emph{arXiv preprint arXiv:2305.14019}, 2023.

\bibitem{comanici2025gemini25pushingfrontier}
\BIBentryALTinterwordspacing
G.~Comanici, E.~Bieber, M.~Schaekermann \emph{et~al.}, ``Gemini 2.5: Pushing the frontier with advanced reasoning, multimodality, long context, and next generation agentic capabilities,'' 2025. [Online]. Available: \url{https://arxiv.org/abs/2507.06261}
\BIBentrySTDinterwordspacing

\bibitem{englhardt2023exploringcharacterizinglargelanguage}
\BIBentryALTinterwordspacing
Z.~Englhardt, R.~Li, D.~Nissanka, Z.~Zhang, G.~Narayanswamy, J.~Breda, X.~Liu, S.~Patel, and V.~Iyer, ``Exploring and characterizing large language models for embedded system development and debugging,'' 2023. [Online]. Available: \url{https://arxiv.org/abs/2307.03817}
\BIBentrySTDinterwordspacing

\bibitem{Assertllm}
\BIBentryALTinterwordspacing
W.~Fang, M.~Li, M.~Li, Z.~Yan, S.~Liu, Z.~Xie, and H.~Zhang, ``Assertllm: Generating and evaluating hardware verification assertions from design specifications via multi-llms,'' 2024. [Online]. Available: \url{https://arxiv.org/abs/2402.00386}
\BIBentrySTDinterwordspacing

\bibitem{gadde2024all}
D.~N. Gadde, A.~Kumar, T.~Nalapat, E.~Rezunov, and F.~Cappellini, ``All artificial, less intelligence: Genai through the lens of formal verification,'' \emph{arXiv preprint arXiv:2403.16750}, 2024.

\bibitem{garcia2025turtle}
D.~Garcia-Gasulla, G.~Kestor, E.~Parisi, M.~Albert{\'\i}-Binimelis, C.~Gutierrez, R.~M. Ghorab, O.~Montenegro, B.~Homs, and M.~Moreto, ``Turtle: A unified evaluation of llms for rtl generation,'' \emph{arXiv preprint arXiv:2504.01986}, 2025.

\bibitem{huang2024towards}
H.~Huang, Z.~Lin, Z.~Wang, X.~Chen, K.~Ding, and J.~Zhao, ``Towards llm-powered verilog rtl assistant: Self-verification and self-correction,'' \emph{arXiv preprint arXiv:2406.00115}, 2024.

\bibitem{huang2024opencoder}
S.~Huang, T.~Cheng, J.~K. Liu, J.~Hao, L.~Song, Y.~Xu, J.~Yang, J.~Liu, C.~Zhang, L.~Chai \emph{et~al.}, ``Opencoder: The open cookbook for top-tier code large language models,'' \emph{arXiv preprint arXiv:2411.04905}, 2024.

\bibitem{hui2024qwen2}
B.~Hui, J.~Yang, Z.~Cui, J.~Yang, D.~Liu, L.~Zhang, T.~Liu, J.~Zhang, B.~Yu, K.~Lu \emph{et~al.}, ``Qwen2. 5-coder technical report,'' \emph{arXiv preprint arXiv:2409.12186}, 2024.

\bibitem{jha2024automated}
S.~K. Jha, S.~Jha, M.~R.~H. Rashed, R.~Ewetz, and A.~Velasquez, ``Automated synthesis of hardware designs using symbolic feedback and grammar-constrained decoding in large language models,'' in \emph{NAECON 2024-IEEE National Aerospace and Electronics Conference}.\hskip 1em plus 0.5em minus 0.4em\relax IEEE, 2024, pp. 95--100.

\bibitem{jiang2024survey}
J.~Jiang, F.~Wang, J.~Shen, S.~Kim, and S.~Kim, ``A survey on large language models for code generation,'' \emph{arXiv preprint arXiv:2406.00515}, 2024.

\bibitem{jimenez2023swe}
C.~E. Jimenez, J.~Yang, A.~Wettig, S.~Yao, K.~Pei, O.~Press, and K.~Narasimhan, ``Swe-bench: Can language models resolve real-world github issues?'' \emph{ICLR}, 2024.

\bibitem{10691770}
F.~R. Kashanaki, M.~Zakharov, and J.~Renau, ``Hdleval benchmarking llms for multiple hdls,'' in \emph{2024 IEEE LLM Aided Design Workshop (LAD)}, 2024, pp. 1--5.

\bibitem{uart}
Kelvin, ``Uart communication protocol and how it works,'' \url{https://www.seeedstudio.com/blog/2022/09/08/uart-communication-protocol-and-how-it-works/}, 2022, accessed: 16th April 2025.

\bibitem{kumari2017interfacing}
R.~S.~S. Kumari and C.~Gayathri, ``Interfacing of mems motion sensor with fpga using i2c protocol,'' in \emph{2017 International Conference on Innovations in Information, Embedded and Communication Systems (ICIIECS)}.\hskip 1em plus 0.5em minus 0.4em\relax IEEE, 2017, pp. 1--5.

\bibitem{li2025autosilicon}
C.~Li, C.~Chen, Y.~Pan, W.~Xu, Y.~Liu, K.~Chang, Y.~Wang, M.~Wang, Y.~Wang, H.~Li \emph{et~al.}, ``Autosilicon: Scaling up rtl design generation capability of large language models,'' \emph{ACM Transactions on Design Automation of Electronic Systems}, 2025.

\bibitem{li2023starcoder}
R.~Li, L.~B. Allal, Y.~Zi, N.~Muennighoff, D.~Kocetkov, C.~Mou, M.~Marone, C.~Akiki, J.~Li, J.~Chim \emph{et~al.}, ``Starcoder: may the source be with you!'' \emph{arXiv preprint arXiv:2305.06161}, 2023.

\bibitem{liu2023verilogeval}
M.~Liu, N.~Pinckney, B.~Khailany, and H.~Ren, ``Verilogeval: Evaluating large language models for verilog code generation,'' in \emph{2023 IEEE/ACM International Conference on Computer Aided Design (ICCAD)}.\hskip 1em plus 0.5em minus 0.4em\relax IEEE, 2023, pp. 1--8.

\bibitem{lozhkov2024starcoder}
A.~Lozhkov, R.~Li, L.~B. Allal, F.~Cassano, J.~Lamy-Poirier, N.~Tazi, A.~Tang, D.~Pykhtar, J.~Liu, Y.~Wei \emph{et~al.}, ``Starcoder 2 and the stack v2: The next generation,'' \emph{arXiv preprint arXiv:2402.19173}, 2024.

\bibitem{RTLLM}
\BIBentryALTinterwordspacing
Y.~Lu, S.~Liu, Q.~Zhang, and Z.~Xie, ``Rtllm: An open-source benchmark for design rtl generation with large language model,'' 2023. [Online]. Available: \url{https://arxiv.org/abs/2308.05345}
\BIBentrySTDinterwordspacing

\bibitem{lu2024rtllm}
------, ``Rtllm: An open-source benchmark for design rtl generation with large language model,'' in \emph{2024 29th Asia and South Pacific Design Automation Conference (ASP-DAC)}.\hskip 1em plus 0.5em minus 0.4em\relax IEEE, 2024, pp. 722--727.

\bibitem{spi}
MikeGrusin, ``Serial peripheral interface (spi),'' \url{https://learn.sparkfun.com/tutorials/serial-peripheral-interface-spi/all}, accessed: 16th April 2025.

\bibitem{achiam2023gpt}
OpenAI, ``Gpt-4.1,'' 2025.

\bibitem{openai2024gpt4technicalreport}
\BIBentryALTinterwordspacing
e.~a. OpenAI, ``Gpt-4 technical report,'' 2024. [Online]. Available: \url{https://arxiv.org/abs/2303.08774}
\BIBentrySTDinterwordspacing

\bibitem{pan2025survey}
J.~Pan, G.~Zhou, C.-C. Chang, I.~Jacobson, J.~Hu, and Y.~Chen, ``A survey of research in large language models for electronic design automation,'' \emph{ACM Transactions on Design Automation of Electronic Systems}, 2025.

\bibitem{subero2024usart}
A.~Subero, ``Usart, spi, i2c, and communication protocols,'' in \emph{Programming PIC Microcontrollers with XC8: Mastering Classical Embedded Design}.\hskip 1em plus 0.5em minus 0.4em\relax Springer, 2024, pp. 297--366.

\bibitem{swaroopa2024evaluating}
S.~Swaroopa, R.~Mukherjee, A.~Debnath, and R.~S. Chakraborty, ``Evaluating large language models for automatic register transfer logic generation via high-level synthesis,'' \emph{arXiv preprint arXiv:2408.02793}, 2024.

\bibitem{thakur2023benchmarking}
S.~Thakur, B.~Ahmad, Z.~Fan, H.~Pearce, B.~Tan, R.~Karri, B.~Dolan-Gavitt, and S.~Garg, ``Benchmarking large language models for automated verilog rtl code generation,'' in \emph{2023 Design, Automation \& Test in Europe Conference \& Exhibition (DATE)}.\hskip 1em plus 0.5em minus 0.4em\relax IEEE, 2023, pp. 1--6.

\bibitem{thakur2024verigen}
S.~Thakur, B.~Ahmad, H.~Pearce, B.~Tan, B.~Dolan-Gavitt, R.~Karri, and S.~Garg, ``Verigen: A large language model for verilog code generation,'' \emph{ACM Transactions on Design Automation of Electronic Systems}, vol.~29, no.~3, pp. 1--31, 2024.

\bibitem{vijayaraghavan2024vhdlevalframeworkevaluatinglarge}
\BIBentryALTinterwordspacing
P.~Vijayaraghavan, L.~Shi, S.~Ambrogio, C.~Mackin, A.~Nitsure, D.~Beymer, and E.~Degan, ``Vhdl-eval: A framework for evaluating large language models in vhdl code generation,'' 2024. [Online]. Available: \url{https://arxiv.org/abs/2406.04379}
\BIBentrySTDinterwordspacing

\bibitem{i2c}
P.~Waite, ``Understanding the i2c bus: A beginner's guide to simplifying communication,'' \url{https://wraycastle.com/blogs/knowledge-base/understanding-the-i2c-bus-a-beginners-guide-to-simplifying-communication }, 2024, accessed: 16th April 2025.

\bibitem{wei2025vflow}
Y.~Wei, Z.~Huang, H.~Li, W.~W. Xing, T.-J. Lin, and L.~He, ``Vflow: Discovering optimal agentic workflows for verilog generation,'' \emph{arXiv preprint arXiv:2504.03723}, 2025.

\bibitem{yubeaton2025verithoughts}
P.~Yubeaton, A.~Nakkab, W.~Xiao, L.~Collini, R.~Karri, C.~Hegde, and S.~Garg, ``Verithoughts: Enabling automated verilog code generation using reasoning and formal verification,'' \emph{arXiv preprint arXiv:2505.20302}, 2025.

\end{thebibliography}
%%%%%%%%%%%%%%%%%%%%%%%%%%%%%%%%%%%%

\end{document}